\documentstyle[axodraw,colordvi]{frank}

\def\beq{\begin{equation}}
\def\eeq{\end{equation}}
\def\bea{\begin{eqnarray}}
\def\eea{\end{eqnarray}}
\def\barr{\begin{array}}
\def\earr{\end{array}}
\newcommand{\sm}{standard model}
\newcommand{\cm}{centre of mass}
\newcommand{\xs}{cross section}
\newcommand{\dxs}{differential cross section}
\newcommand{\lc}{linear collider}
\newcommand{\pe}{\mbox{$e^+e^-$}}
\newcommand{\ee}{\mbox{$e^-e^-$}}
\newcommand{\ep}{\mbox{$e^-\gamma$}}
\newcommand{\pp}{\mbox{$\gamma\gamma$}}

\newcommand{\bil}{bilepton}

\newcommand{\guts}{grand unified theories}

\newcommand{\kst}{Kolmogorov-Smirnov test}
\newcommand{\isr}{initial state radiation}
\newcommand{\bs}{beamstrahlung}

\newcommand{\EM}{electromagnetic}
\def\lsim{\mathrel{\vcenter{\hbox{$<$}\nointerlineskip\hbox{$\sim$}}}}
\def\gsim{\mathrel{\vcenter{\hbox{$>$}\nointerlineskip\hbox{$\sim$}}}}

\begin{document}

\begin{flushright}
PSI-PR-97-11\\
FTUV/97-13\\
IFIC/97-13\\
March 1997
\end{flushright}

\vfill

\begin{frontmatter}
\title{Discovering and Studying Bileptons \\
	with \ee\ Collisions}
\author{Frank Cuypers}
\address{{\tt cuypers@psi.ch}\\
        Paul Scherrer Institute,
        CH-5232 Villigen PSI,
        Switzerland}
\author{Martti Raidal}
\address{{\tt raidal@titan.ific.uv.es}\\
        Department of Theoretical Physics and IFIC, \\
        University of Valencia,
        46100 Burjassot, Valencia, Spain}
\begin{abstract}
We analyze the prospects 
for discovering and unraveling the nature
of doubly-charged \bil s
at a \lc\ of the next generation
running in its \ee\ mode.
We stress the importance of \isr, beam spread and polarization,
and compute the discovery bounds.
The gauge nature of vector \bil s
can be determined
by studying hard photon emission.
\end{abstract}

\begin{keyword}
        bileptons,
        dileptons,
        new physics,
        new bosons,
	\ee\ collisions
\PACS
        14.80.-j, 
        12.60.-i, 
        12.10.Dm, 
        29.17.+w
\end{keyword}

\end{frontmatter}

\vfill
\clearpage

\section{Introduction}

While the LHC offers an entry 
into the the high energy regime of the \sm\
with a significant opportunity for discovering new phenomena,
a linear electron collider of the next generation \cite{lc}
will provide a complementary program of experiments
with unique opportunities 
for both discoveries and precision measurements.
A major asset to fulfill this purpose
is the versatility of \lc s,
as they can be operated in the four \pe, \ee, \ep\ and \pp\ modes,
with highly polarized electron and photon beams.
Moreover,
starting from a \cm\ energy of several hundred GeV,
it will later be possible to upgrade these machines
into the TeV range.

The \ee\ running mode~\cite{e-e-}
is a particularly interesting feature of a high energy \lc.
One of the many promising processes 
which can be studied in this mode 
is the resonant production of doubly-charged \bil s.
These particles are predicted 
by many extensions of the \sm,
such as \guts~\cite{guts},
theories with enlarged Higgs sectors~\cite{higgs},
technicolour theories~\cite{tc},
theories of compositeness~\cite{comp}
and theories which generate neutrino majorana masses~\cite{maj}.
They may appear as scalars, vectors or Yang-Mills fields
and can be classified in a model-independent way~\cite{cd}
much like leptoquarks~\cite{brw}.

The prospects for discovering doubly-charged \bil s
through their $s$-channel resonance
in \ee\ collisions
have been considered previously 
within the framework of specific models~\cite{ee2bil,jack}\footnote{
We do not agree with the \xs s (3.30) and (1)
stated in the first two references of Ref.~\cite{ee2bil},
respectively.
}.
We analyze these reactions here in more details
taking into account properties of the beams 
and effects of initial state radiation.
We also study in a model-independent way 
the range of bilepton couplings and masses 
which can be probed in the \ee\ \lc\ mode
and we present ways of disentangling 
the different types of \bil s
from each other.

For this we present in the following section
the most general classification of bileptons 
together with their interaction lagrangians
with fermions and gauge fields.
In Section 3 
we study the discovery potential of \bil s
and we demonstrate that the radiative return to resonance 
due to \isr\ 
may enable the discovery of \bil s
without having to resort to a painstaking scan of \cm\ energies.  
In Section 4 
we investigate how to distinguish between scalar and vector \bil s,
and to which extent a determination of the gauge nature of vector \bil s
is possible.
We conclude in Section 6.

\section{Bilepton Classification}

We define \bil s to be bosons 
which couple to two leptons
and which carry two units of lepton number.
Their interactions need not necessarily conserve lepton flavour,
but otherwise we demand the symmetries of the \sm\ to be respected.
The most general renormalisable lagrangian
of this kind
is of dimension four
and involves seven \bil\ fields
$L_1^{+},\ \tilde L_1^{++},\ L_{2\mu}^{+},\ L_{2\mu}^{++},\
L_{3}^{0},\ L_{3}^{+}$ and $L_{3}^{++}$.
It is given by~\cite{cd}

\bea
\label{lag}
{\cal L}
&~=~&
-~
\lambda_1^{ij} \quad L_1^{+} \quad 
\left( \bar \ell^c_i P_L \nu_j - \bar \ell^c_j P_L \nu_i \right) \\\nonumber
&&
+~
\tilde\lambda_1^{{ij}} \quad \tilde L_1^{++} 
\quad \bar \ell^c_i P_R \ell_j \\\nonumber
&&
+~
\lambda_2^{ij} \quad L_{2\mu}^{+} 
\quad \bar\nu^c_i \gamma^\mu P_R \ell_j \\\nonumber
&&
+~
\lambda_2^{ij} \quad L_{2\mu}^{++} 
\quad \bar \ell^c_i \gamma^\mu P_R \ell_j \\\nonumber
&&
+~
\sqrt{2}\lambda_3^{{ij}} \quad L_{3}^{0} 
\quad \bar\nu^c_i P_L \nu_j \\\nonumber
&&
-~
\lambda_3^{{ij}} \quad L_{3}^{+} 
\quad \left( \bar \ell^c_i P_L \nu_j + \bar \ell^c_j P_L \nu_i \right) \\\nonumber
&&
-~
\sqrt{2}\lambda_3^{{ij}} \quad L_{3}^{++} 
\quad \bar \ell^c_i P_L \ell_j \\\nonumber
&&
+~
\mbox{ h.c.}~
~,
\eea

where the subscripts 1--3 label the dimension
of the $SU(2)_L$ representation
to which the \bil s belong
and the indices $i,j=e,\mu,\tau$ stand for the lepton flavours.
The chirality projection operators
are defined as 
$P_{R,L} = (1\pm\gamma_5)/2$.

In the following 
we shall concentrate on the doubly-charged members
of the scalar singlet $\tilde L_1^{--}$,
the vector doublet $L_{2\mu}^{--}$
and the scalar triplet $L_3^{--}$.
The coupling matrices of the two types of scalars 
to different flavours of leptons
$\lambda^{ij}$
are hermitian,
while the coupling matrices for vectors 
are arbitrary.
If $CPT$ symmetry is imposed,
the latter are either hermitian or anti-hermitian.

For the sake of definiteness
we shall focus from now on
the universal diagonal coupling matrix

\renewcommand{\arraystretch}{.5}
\beq
\label{univ}
\lambda^{ij}~
\equiv~
\lambda~
\left(\begin{array}{c@{~}c@{~}c}1&&\\&1&\\&&1\end{array}\right)~
~,
\eeq
\renewcommand{\arraystretch}{1}

which would be a natural possibility
for gauge \bil s.
It is straightforward to implement a different choice,
which we will do at the end of this section.
Within the framework of the coupling matrix~(\ref{univ}),
though,
the negative results 
of muonium-antimuonium conversion searches~\cite{willy}
imply approximately the same upper limit
for the three doubly-charged \bil s
on the ratio $\lambda/m_B$,
where $m_B$ is the \bil\ mass:

\beq
\label{muconv}
\lambda~
\lsim~
0.5~
m_B/\mbox{TeV}~
~.
\eeq

Up to now 
we have made no difference
between ordinary vectors
and Yang-Mills fields
for the doublet $L_{2\mu}$ \bil s.
This distinction arises,
however,
when the interactions with the gauge fields are considered.
In particular, the vector \bil\ coupling to photon is 
described by the following lagrangian:

\bea
\label{lagvec}
{\cal L}
& ~=~ & 
-1/2 ~{\left(D_\mu{L_{2\nu}}-D_\nu{L_{2\mu}}\right)}^\dagger ~
           \left(D_\mu{L_{2\nu}}-D_\nu{L_{2\mu}}\right)
\\\nonumber
&&-~
i~
\kappa~
eQ ~{L^\dagger_\mu}{L_{2\nu}} ~
\left(\partial_\mu A_\nu - \partial_\nu A_\mu\right)
~,
\eea

where the covariant derivative is given by

\beq
\label{covder}
D_\mu ~=~ \partial_\mu - ie Q A_\mu
~,
\eeq

and $Q$ is the charge of the \bil,
{\em i.e.},
$Q=2$ for our purposes.
If the vector \bil s are Yang-Mills fields,
the parameter  
$\kappa$
in the second line of Eq.~(\ref{lagvec})
takes the value one at tree level.
On the other hand,
for non-gauge vectors
one expects a {\em minimal coupling} to the photon
which is obtained when
$\kappa$
vanishes.

\section{Discovering Bileptons}

\begin{figure}[htb]
\unitlength.5mm
\SetScale{1.418}
\begin{boldmath}
\begin{center}
\begin{picture}(70,40)(0,0)
\ArrowLine(0,0)(15,15)
\ArrowLine(0,30)(15,15)
\DashLine(15,15)(45,15){1}
\ArrowLine(45,15)(60,30)
\ArrowLine(45,15)(60,0)
\Text(-2,0)[r]{$e^-$}
\Text(-2,30)[r]{$e^-$}
\Text(62,0)[l]{$\mu^-$}
\Text(62,30)[l]{$\mu^-$}
\Text(30,23)[c]{$L^{--}$}
\end{picture}
\end{center}
\end{boldmath}
\bigskip
\caption{
  Lowest order Feynman diagram 
  inducing the process $\ee\to\mu^-\mu^-$.
  The exchanged doubly-charged \bil\ $L^{--}$ 
  can one of the scalars $\tilde L^{--}_1,L^{--}_3$ 
  or the vector $L^{--}_{2\mu}$.
}
\bigskip
\label{fee2mm}
\end{figure}
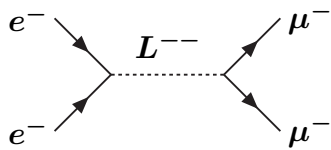

Muons can be pair-produced in \ee\ collisions
via the $s$-channel exchange of a doubly-charged scalar or vector \bil,
as depicted in Fig.~\ref{fee2mm}.
The differential \xs s of these reactions are given by

\bea
\label{diffxs1}
{d\sigma(\tilde L_1^{--}) \over dt}
&~=~&
\underbrace{{(1+P_1)(1+P_2) \over 4}}_{\displaystyle[RR]}~
{\lambda^4 \over 2\pi}~
{1 \over (s-m_B^2)^2 + m_B^2\Gamma_B^2}~~,
\\\nonumber\\
\label{diffxs2}
{d\sigma(L_{2\mu}^{--}) \over dt}
&~=~&
\underbrace{{1-P_1P_2 \over 2}}_{\displaystyle[LR]}~
{\lambda^4 \over 8\pi s^2}~
{t^2 + u^2 \over (s-m_B^2)^2 + m_B^2\Gamma_B^2}~~,
\\\nonumber\\
\label{diffxs3}
{d\sigma(L_3^{--}) \over dt}
&~=~&
\underbrace{{(1-P_1)(1-P_2) \over 4}}_{\displaystyle[LL]}~
{2\lambda^4 \over \pi}~
{1 \over (s-m_B^2)^2 + m_B^2\Gamma_B^2}~
~,
\eea

where 
$P_1$ and $P_2$ are the polarizations of the incoming electron beams,
$m_B$ and $\Gamma_B$ are respectively the mass and width of the \bil,
and $\lambda$ is the universal diagonal coupling (\ref{univ}).
With this assumption on the structure of the leptonic couplings,
the leptonic widths of the \bil s 
are given by~\cite{ee2bil,bildec}\footnote{
We do not agree with the width 
stated in the text of the second reference of Ref.~\cite{ee2bil}.
}

\beq
\label{width}
\Gamma_B~
=~
G~
{\lambda^2 \over 8\pi}~
m_B~,
\qquad
\left\{
\begin{array}{lcl}
\tilde L_1 &:& G = 3~, \\
L_{2\mu} &:& G = 1~, \\
L_3 &:& G = 6~.
\end{array}
\right.
\eeq

Taking into account the current limits~(\ref{muconv}) 
on the possible values of the coupling $\lambda$,
we see that \bil s as light as 200 GeV
are bound to have a decay width
which is significantly less than 1 GeV.

\bigskip

In the case of the doublet and triplet \bil s,
the different members of the multiplet need not necessarily 
have the same mass.
If the doubly-charged members of the multiplets
turn out to be heavier 
than the singly-charged ones,
the non-leptonic decay mode
$L^{--} \to L^-W^-$
can possibly also contribute to the total width~\cite{jack}.
However,
only heavier \bil s
can realistically accommodate a mass splitting 
exceeding the mass of the $W$ boson.
We therefore ignore this possibility here
and assume purely leptonic decays of the \bil s.

Upon integrating Eqs~(\ref{diffxs1}--\ref{diffxs3})
the total \xs s are

\beq
\label{totxs}
\sigma~
=~
S~
{\lambda^4 \over 12\pi}~
{s \over (s-m_B^2)^2 + m_B^2\Gamma_B^2}~,
\qquad
\left\{
\begin{array}{lclrl}
\tilde L_1 &:& S = &6& [RR]~, \\
L_{2\mu} &:& S = &1& [LR]~, \\
L_3 &:& S = &24& [LL]~.
\end{array}
\right.
\eeq

If the decay width of the \bil\ is significantly larger 
than the beam spread,
it is a good approximation to assume monochromatic electron beams.
In this case,
the centre of the resonance can be precisely met,
and the replacement
$s=m_B^2$
into Eq.~(\ref{totxs})
yields the \xs\

\beq
\label{resxs}
\sigma~
=~
{S \over G}~
{16\pi \over 3m_B^2}
~.
\eeq

This \xs\ does not depend on the coupling to leptons,
since any decrease/increase of the strength of the interactions
also results into a decrease/increase of the decay width.
In the observed \xs\ both effects result in a cancellation.
Obviously,
this result stops making sense
for infinitesimal values of the coupling.
Indeed,
when the decay width is smaller 
than the beam spread,
the latter needs to be taken in account~\cite{jack}.
If we approximate it
by a box\footnote{
In reality it resembles more a saddle 
than a box~\cite{zdr}.
}
of width $R$
centrered around the resonance,
the observed \xs\ results from the convolution

\beq
\label{spreadxs}
\sigma~
=~
\int_{m_B-R/2}^{m_B+R/2} d\sqrt{s}~
{1 \over R}~
\sigma(\sqrt{s})~
\simeq~
{S \over G}~
{\pi\lambda^2 \over 3 m_B R}~
\qquad
(\Gamma_B \ll R \ll m_B)~
~,
\eeq

where a typical value for the beam spread is~\cite{zdr}

\beq
\label{spread}
R~=~10^{-2}~\sqrt{s}
~.
\eeq

Note that at a muon collider
the corresponding spread 
would be of the order of $4\cdot 10^{-4}~\sqrt{s}$~\cite{mucol}.
From this fact alone,
plus the luminosity enhancement at nominal energy 
associated with the much reduced Brems- and beamstrahlung rates
of colliding muon beams,
results the possibility of generating 
hundred times more doubly-charged \bil s
at a $\mu^+\mu^-$ facility.

It may be that the resonance is so narrow
that no significant tail 
extends on either side of the resonance.
If this is the case
and if the \cm\ energy were strictly confined within $\sqrt{s} \pm R$,
it would be a long unpleasant task 
to find this resonance in the first place,
through a minute energy scan.
However,
in practice
\isr\ and \bs\
will unavoidably perform a scan at lower energies.
To obtain a rough idea of the effect,
we approximate the \cm\ energy spectrum
after \isr\
to be of the form

\beq
\label{isr}
F(x,s)~
=~
{3\alpha\over\pi}~
\ln{s \over m_e^2}~
{1 \over 1-x}~
~,
\eeq

where 
$\alpha$ is the \EM\ coupling strength,
$m_e$ is the mass of the electron
and 
$x=\hat s/s$ is the squared \cm\ energy fraction 
of the event.
(The expression (\ref{isr}) 
can be derived, {\it e.g.,} 
from the $\ee \to L^{--}\gamma$ \xs\ (\ref{xstot}),
by singling out the large logarithm
in the limit where $m_B^2 \ll s$.)
Upon smearing the \xs\ (\ref{totxs}) with the spectrum (\ref{isr}),
we have

\bea
\label{isrxs}
\sigma~
&=~&
\int_0^1 dx~F(x,s)~\sigma(xs)~  \nonumber \\
&\simeq~&
{S \over G}~
2\alpha\lambda^2~
\ln{s \over m_e^2}~
{1 \over s - m_B^2}~
\qquad
(\Gamma_B \ll m_B \ll \sqrt{s})~
~.
\eea

To estimate the discovery potential,
we use the following scaling relation for the \ee\ luminosity

\beq
\label{lum}
{\cal L}_{e^-e^-}~
=~
3.25\cdot 10^7~s~
~,
\eeq

which closely corresponds 
to a luminosity of 25 fb$^{-1}$ at $\sqrt{s}=500$ GeV
which scales like the square of the \cm\ energy.
This choice for the luminosity
is dictated by the latest \pe\ \lc\ design reports~\cite{zdr}
and the fact that the \ee\ mode 
will approximately suffer a 50\%\ luminosity reduction
because of the anti-pinch effect \cite{jim}.

If we assume that observing one flavour violating $\ee\to\mu^-\mu^-$ event
already constitutes a discovery,
we need an average number of 
$-\ln(1-p)$
Poisson distributed events 
such that {\em at least} 1 event
is observed with probability $p$.
Hence,
a predicted average of at least 3 events
is needed to guarantee a discovery with 95\%\ confidence.

These considerations
imply that the minimal value of the lepton-\bil\ coupling $\lambda$
needed to observe at least one $\ee\to\mu^-\mu^-$ event
at the 95\% confidence level
on the resonance
is
according to Eqs~(\ref{spreadxs},\ref{spread})

\beq
\label{limres}
\lambda~
\gsim~
10^{-4}~
\sqrt{ {G \over S}~{6 \over 25\pi} }~
\simeq~
{1\over4}~
10^{-4}~
\sqrt{G \over S}~
~.
\eeq

Similarly,
according to the \xs\ (\ref{isrxs}),
far off the resonance 
the minimal coupling
is

\beq
\label{limisr}
\lambda~
\gsim~
10^{-4}~
\sqrt{ {G \over S}~{4 \over \alpha \ln{s \over m_e^2}}~{s-m_B^2 \over s} }~
\simeq~
{1\over2}~
10^{-3}~
\sqrt{G \over S}~
~.
\eeq

Of course,
the \isr\ spectrum (\ref{isr})
is only an approximation,
which at least has the virtue
of yielding qualitative analytic results.
We now present
the results of a more precise
numerical convolution
with a more realistic resummed luminosity distribution~\cite{kf}.
We also assume a polarization of 90\%\
for the incoming electron beams.

\bigskip
\begin{figure}[htb]
\centerline{\input{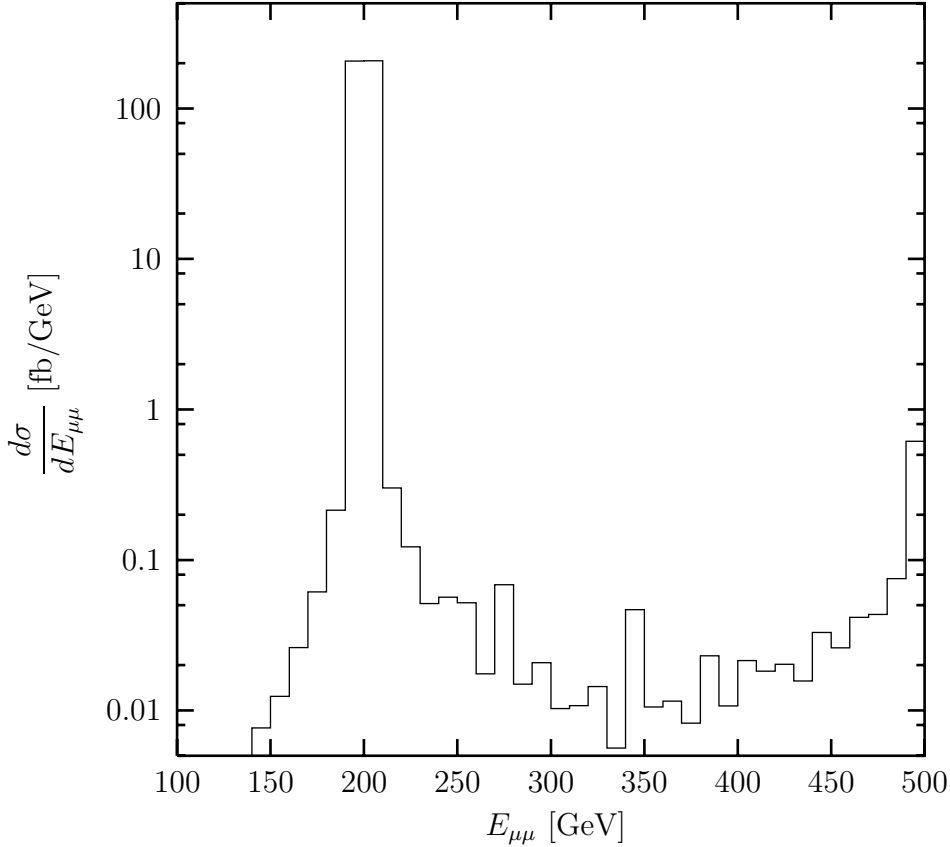}}
\bigskip
\caption{
Energy distribution of the muon pairs
in the reaction $\ee\to\mu^-\mu^-$
at the \cm\ energy $\protect\sqrt{s}=500$ GeV.
This processes is mediated 
by a doubly-charged vector \bil\ $L_{2\mu}^{--}$
of mass $m_B=200$ GeV
which couples with strength $\lambda=0.1$
to the leptons (\protect\ref{univ}).
The peak at $E_{\mu\mu}=m_B=200$ GeV
results from the radiative return
to the \bil\ resonance
due to \isr.
}
\bigskip
\label{ffull}
\end{figure}

\begin{figure}[htb]
\centerline{\input{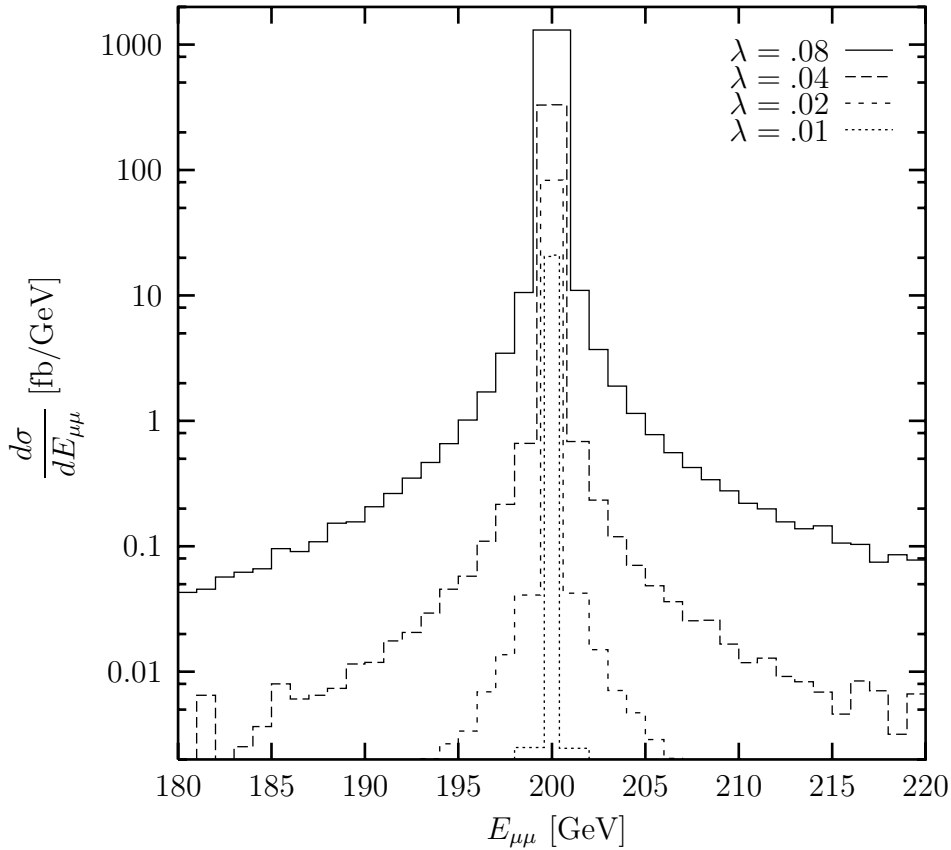}}
\bigskip
\caption{
Same as Fig.~\protect\ref{ffull}
for different values of the lepton-\bil\ coupling $\lambda$
and focusing on the muon energy region located around the peak.
For the purpose of a clear display
the two centremost bins for smaller couplings are depicted narrower 
than in the simulation.
Their actual size is 1 GeV,
like all other bins.
}
\bigskip
\label{fzoom}
\end{figure}

In Fig.~\ref{ffull}
we show 
the energy distribution of the muon pairs
over its full range.
The peak at 200 GeV
corresponds to the mass of a vector \bil\ $L_{2\mu}^{--}$
exchanged in the $s$-channel.
The coupling to leptons is set to $\lambda=0.1$.
Fig.~\ref{fzoom}
displays a zoom into this peak region
for different values of the coupling.
The effect of the radiative return to the resonance is important
even for bilepton masses far below the nominal collider energy.
This suggests that no painstaking energy scan may be needed
to localize a possible narrow \bil\ resonance. 

\bigskip

\begin{figure}[htb]
\centerline{\input{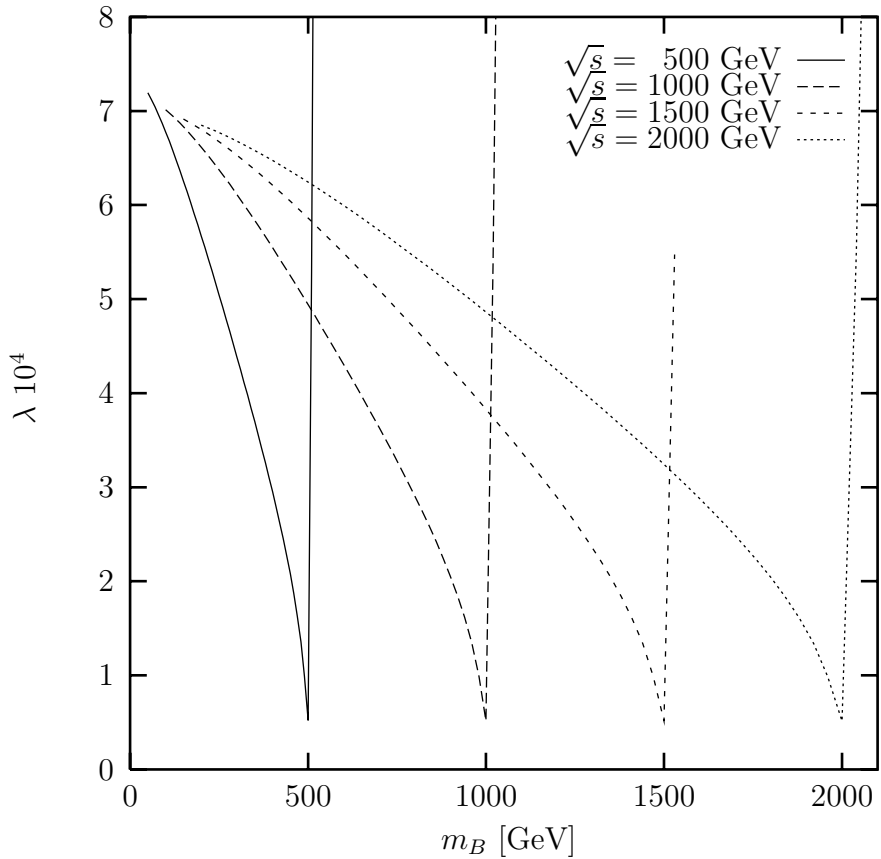}}
\bigskip
\caption{
Discovery prospects of a doubly-charged vector \bil\ $L_{2\mu}^{--}$
as a function of its mass $m_B$ 
and coupling to leptons $\lambda$ (\protect\ref{univ})
for several values of the cm\ energy $\protect\sqrt{s}$.
In the absence of any $\ee\to\mu^-\mu^-$ event,
the areas below and to the right of the curves 
cannot be excluded 
to better than 95\% confidence.
}
\bigskip
\label{fdiscovery}
\end{figure}

In Fig.~\ref{fdiscovery}
we show for several typical \cm\ energies
the curves in the $(m_B,\lambda)$ parameter space
which correspond to having a 95\% probability 
of seeing a $\ee\to\mu^-\mu^-$ event
due to the $s$-channel exchange of a vector \bil\ $L_{2\mu}^{--}$.
As expected
the sensitivity is maximal 
when $m_B=\sqrt{s}$
and worsens dramatically below threshold.
Beyond threshold,
however,
the sensitivity is at worst decreased by about an order magnitude.

These results turn out to be actually quite close 
to the analytical predictions of Eqs~(\ref{limres},\ref{limisr}).
As expected according to these relations,
the corresponding limits 
for the singlet $\tilde L_1^{--}$
and triplet $L_3^{--}$ \bil s
are shifted down by factors of $\sqrt{2}$ and $2$,
respectively.

\bigskip

While we have chosen the bilepton-lepton coupling matrix 
to be of the form (\ref{univ}), 
off-diagonal matrix elements $\lambda^{ij}$
($i\neq j$)
are in principle also allowed.
For instance,
in the case of scalar bileptons 
$\lambda^{ij}$ is a matrix of Yukawa couplings 
which, in general, need not be flavour diagonal
and would lead to different final states 
in the processes $\ee\to\ell_i^-\ell_j^-$.
In the presence of such non-diagonal couplings 
the quantity plotted in Fig.~\ref{fdiscovery} 
against bilepton mass $m_B$ 
becomes
$\lambda^{ee}\lambda^{ij} / \sqrt{\sum_{kl}(\lambda^{kl})^2/3}$. 
It could thus be 
that the electron-\bil\ coupling $\lambda^{ee}$ is too small 
(much less than $10^{-4}$)
to preclude the discovery of \bil s in \ee\ collisions.
In this case
they can still be pair- or singly-produced at a \lc\
in \pe , \pp\  and \ep\ scattering~\cite{lam}.

The low-energy constraints 
on the off-diagonal couplings~\cite{cd,lnv} 
are generally more stringent than the bound (\ref{muconv}). 
Still,
the combination of all \lc\ operating modes
will eventually probe the \bil s
to exceedingly better sensitivities 
than any of the current experiments.

\section{Determining the Nature of Bileptons}

A standard way of determining the spin of a particle
is to examine the angular distribution of its decay products.
Obviously,
scalar decays must be isotropic
whereas vector decays will display 
some non-trivial angular dependence.
This is clearly verified
by the \dxs s (\ref{diffxs1}--\ref{diffxs3}).

However,
since the availability of high electron beam polarizations
(in excess of 80\%)
is not a matter of debate~\cite{zdr},
the polarization dependence of the scalar and vector total \xs s
(\ref{totxs})
will permit a much more straightforward determination 
of the spin of any discovered \bil.
Indeed,
by flipping the helicity of one of the beams
from the configuration with, say, 
two identical polarizations
($P_1=P_2=P$)
to the one with two opposite polarizations,
($P_1=-P_2=P$)
a scalar signal would be suppressed by a factor
$(1-P^2) / (1+P)^2$,
whereas a vector signal would be enhanced by a factor
$(1+P^2) / (1-P^2)$.
Even with a polarization of only 80\% ($P=0.8$),
this would mean a reduction or increase
of the signal by respectively factors of almost 10 and 5.
There is no way such a dramatic effect would escape detection.

However, determining the gauge nature 
of a discovered doubly-charged vector \bil\ $L_{2\mu}^{--}$
is not such an easy task.
It will necessarily involve
a precise study of the interaction 
between photons and \bil s,
according to Eq.~(\ref{lagvec}).

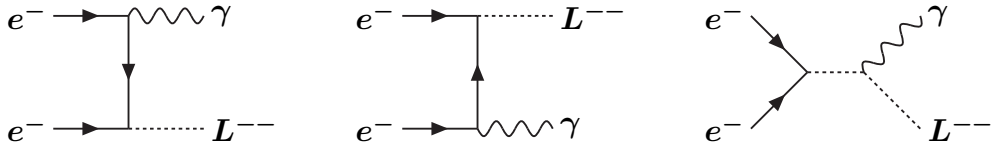
\begin{figure}[htb]
\unitlength.5mm
\SetScale{1.418}
\begin{boldmath}
\begin{center}
\begin{picture}(90,40)(-25,0)
\ArrowLine(00,00)(20,00)
\ArrowLine(00,30)(20,30)
\ArrowLine(20,30)(20,00)
\Photon(20,30)(40,30){2}{3}
\DashLine(20,00)(40,00){1}
\Text(-2,00)[r]{$e^-$}
\Text(-2,30)[r]{$e^-$}
\Text(42,00)[l]{$L^{--}$}
\Text(42,30)[l]{$\gamma$}
\end{picture}
\begin{picture}(90,40)(-25,0)
\ArrowLine(00,00)(20,00)
\ArrowLine(00,30)(20,30)
\ArrowLine(20,00)(20,30)
\Photon(20,00)(40,00){-2}{3}
\DashLine(20,30)(40,30){1}
\Text(-2,00)[r]{$e^-$}
\Text(-2,30)[r]{$e^-$}
\Text(42,30)[l]{$L^{--}$}
\Text(42,00)[l]{$\gamma$}
\end{picture}
\begin{picture}(90,40)(-25,0)
\ArrowLine(00,00)(15,15)
\ArrowLine(00,30)(15,15)
\DashLine(15,15)(30,15){1}
\Photon(30,15)(45,30){2}{3}
\DashLine(30,15)(45,00){1}
\Text(-2,00)[r]{$e^-$}
\Text(-2,30)[r]{$e^-$}
\Text(47,00)[l]{$L^{--}$}
\Text(47,30)[l]{$\gamma$}
\end{picture}
\end{center}
\end{boldmath}
\bigskip
\caption{
  Lowest order Feynman diagrams 
  responsible for the reaction $\ee \to L^{--}_{2\mu}\gamma$.
}
\bigskip
\label{fee2bg}
\end{figure}

We propose here a method based on the reaction

\beq
\label{ee2bg}
\ee~ \to~ L_{2\mu}^{--}\gamma~
~,
\eeq

which proceeds through the three Feynman diagrams
depicted in Fig.~\ref{fee2bg}.
The third of these graphs
involves the \bil-photon coupling.
As before,
the lepton-flavour-violating decays of the produced \bil\
guarantee an unmistakable background-free signal.

\bigskip

The total \xs\ of the process (\ref{ee2bg})
is given by

\bea
\label{xstot}
\sigma~
=~
{\alpha \lambda^2 \over 3(s-m_B^2)}~
&\Bigg\{~&
{1+P^2\over2} ~{3\Lambda^2} 
\\\nonumber
&+~&
{1-P^2\over2}~
\bigg[~
-~ \kappa^2 ~(1+\Lambda-\Lambda^2)
\\\nonumber&&\qquad\qquad\quad
+~ 2\kappa ~(1+\Lambda-2\Lambda^2)
\\\nonumber&&\qquad\qquad\quad
-~ 37 + 35\Lambda - 17\Lambda^2
\\\nonumber&&\qquad\qquad\quad
+~ 6 ( \Lambda^2 - 2\Lambda + 2 ) ~\ln{1+\rho\over1-\rho}~
\\\nonumber&&\qquad\qquad\quad
+~ {s \over m_B^2} ~( 1 - \kappa )^2
\bigg]~
\Bigg\}~
~,
\eea

where

\beq
\Lambda ~=~ 1 - {m_B^2 \over s} 
\qquad\qquad\mbox{and}\qquad\qquad
\rho ~=~ \sqrt{1 - {4m_e^2 \over s}}~
~.
\eeq

Although we have set the electron mass $m_e$ to zero 
wherever it is irrelevant 
in the final \xs\ (\ref{xstot}),
it is important to keep it finite throughout the calculation
in order not to miss those terms
which result from a cancellation of the electron mass
in the numerator and the denominator.
In particular,
the {\em helicity flip} term 
on the first line 
is due to such a cancellation.
As expected for such {\em helicity flip} terms,
it vanishes at threshold
where the photon energy is small.

\bigskip
\begin{figure}[htb]
\centerline{\input{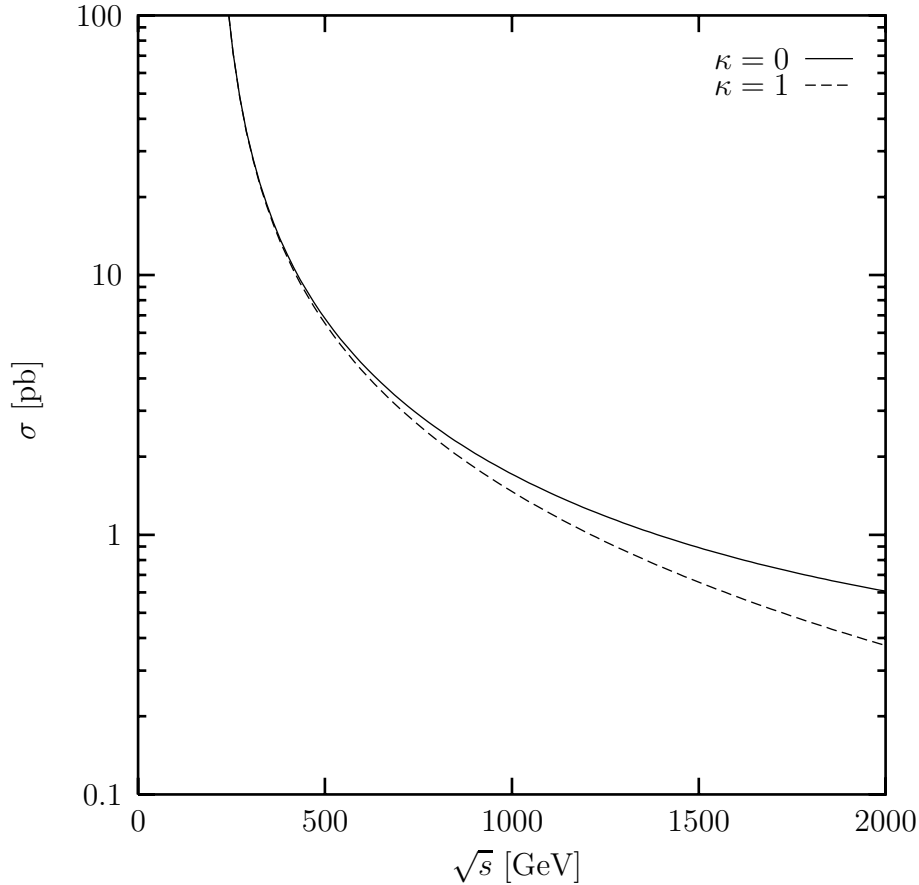}}
\bigskip
\caption{
Cross section for the process $\ee \to L^{--}_{2\mu}\gamma$
as a function of the \cm\ energy.
The mass of the produced \bil\ is 200 GeV
and its coupling to leptons is 0.1.
}
\bigskip
\label{fxs}
\end{figure}

Close to threshold
the \xs\ (\ref{xstot}) is dominated
by the large logarithm in the penultimate term.
Since this logarithm is solely due 
to the $t$- and $u$-channel electron exchanges,
the gauge nature of the \bil\ 
can only be revealed at higher energies.
Away from threshold
the last term,
proportional to $s/m_B^2$,
becomes increasingly more important.
All the other terms 
are numerically irrelevant 
at all energies.

At asymptotic energies
the \xs\ (\ref{xstot}) saturates 
in the case of an ordinary vector
with $\kappa=0$.
In contrast,
if the produced \bil\ is a Yang-Mills field
with $\kappa=1$,
the \xs\ decreases like $1/s$.
Although in principle this phenomenon may be used 
to distinguish these two cases,
the onset of this asymptotic regime
may be too slow 
to provide a useful test 
at the initial operating energies of the \lc.
This is demonstrated in Fig.~\ref{fxs},
where we have plotted the energy dependence of the \xs\ (\ref{xstot}) 
for an ordinary ($\kappa=0$) 
and a Yang-Mills ($\kappa=1$) \bil\ of 200 GeV
coupling with strength $\lambda=0.1$ to leptons.
The difference is even smaller for heavier \bil s.

\bigskip
\begin{figure}[htb]
\centerline{\input{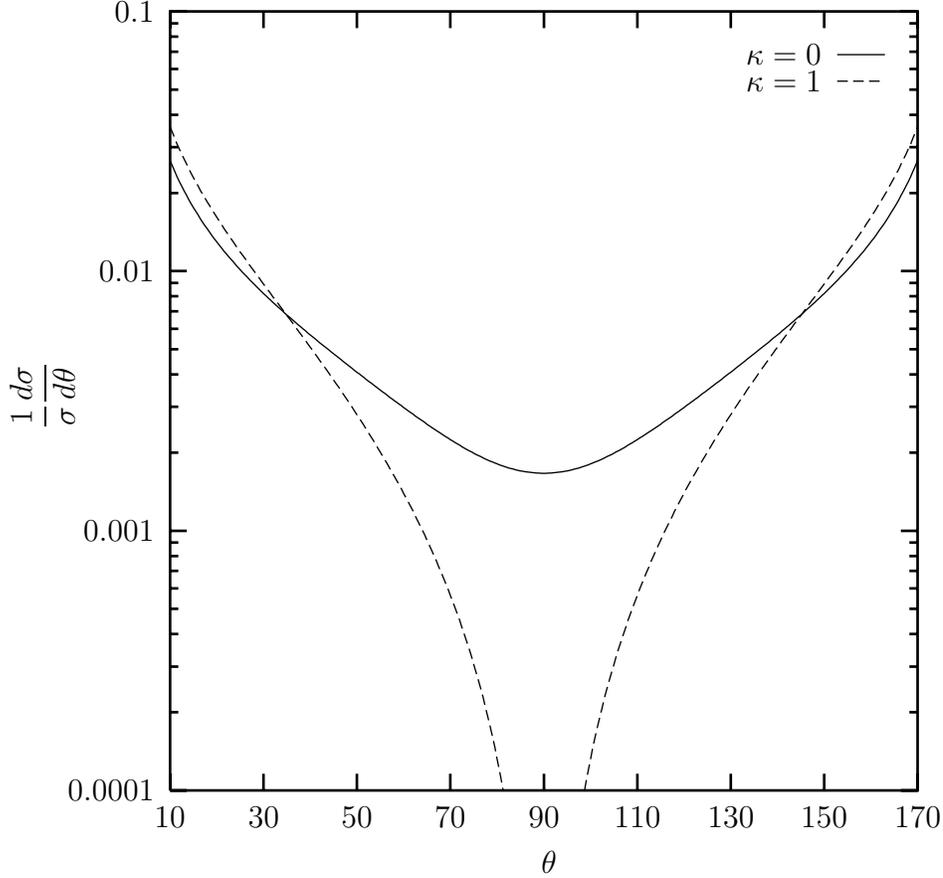}}
\bigskip
\caption{
Angular distribution of the process $\ee \to L^{--}_{2\mu}\gamma$
for a \bil\ of 200 GeV
and a \cm\ energy of 500 GeV.
}
\bigskip
\label{fdxs}
\end{figure}

However,
the gauge nature of the \bil s
may also be determined 
from the angular distribution of the emitted photons.
We do not give here the analytical form of the \dxs\
corresponding to the process (\ref{ee2bg}),
since it is long and not particularly enlightening.
However,
we plot this distribution in Fig.~\ref{fdxs}
for an ordinary and a Yang-Mills vector \bil\
of mass 200 GeV
at a \cm\ energy of 500 GeV.
The most salient feature
is the existence of an radiation amplitude zero
in the transverse direction
for the gauge \bil.

\bigskip
\begin{figure}[htb]
\centerline{\input{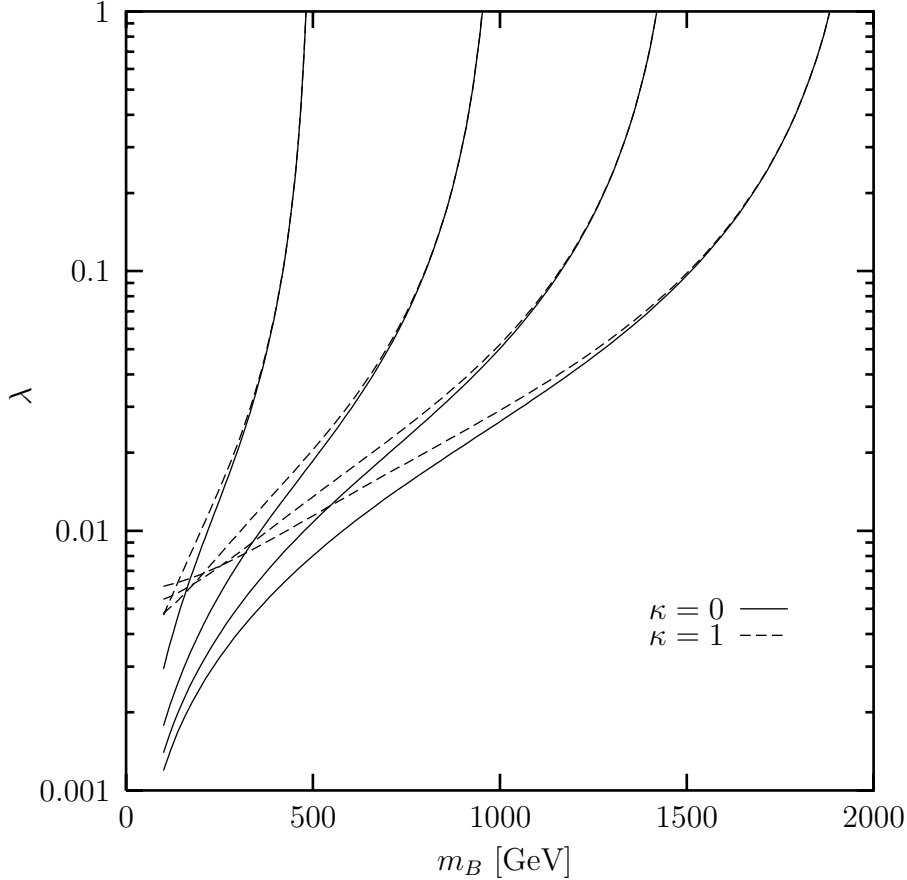}}
\bigskip
\caption{
Prospects for determining the gauge nature 
of a doubly-charged vector \bil\ $L_{2\mu}^{--}$
as a function of its mass $m_B$ 
and coupling to leptons $\lambda$ (\protect\ref{univ}).
The \cm\ energy increases from left to right
as $\protect\sqrt{s}=0.5$, 1, 1.5 and 2 TeV.
The areas below and to the right of the curves 
do not allow a distinction between an ordinary vector
and a Yang-Mills field
to better than 99.9\% confidence.
The plain curves correspond to the case 
where the observed \bil\ is an ordinary vector ($\kappa=0$)
whereas the dashed curved correspond to the case 
where the observed \bil\ is a Yang-Mills field ($\kappa=1$).
}
\bigskip
\label{fks}
\end{figure}

To make efficient use of this difference in the angular distributions
we perform a \kst~\cite{cr}.
For this
we first rewrite the angular distributions
in term of the angular variable 
$y=\cos^2\theta$
in order to make advantageously use 
of the forward-backward symmetry
of this process.
We then integrate these new distributions
from $y=0$ 
up to an arbitrary value of $y$ 
such that $y\le\cos^2\theta_{\rm min}$,
where $\theta_{\rm min}$ is the smallest acceptance of the detector.
This defines two cumulative probabilities
(for $\kappa=0$ and $\kappa=1$)
which increase monotonously 
as a function of $y$
from 0 at $y_{\rm min}=0$
up to 1 at $y_{\rm max}=\cos^2\theta_{\rm min}$.
Between these boundaries
the cumulative probabilities for the two cases are different. 
Somewhere in the middle of the range 
the difference between the cumulative probabilities is maximum. 
The Kolmogorov-Smirnov statistic $D$
is defined 
as the product of this number
times
the square root of the number of observed events.

If we are dealing with, say, a Yang-Mills \bil\ ($\kappa=1$),
this cumulative probability is measured experimentally
whereas the corresponding cumulative probability 
for the ordinary vector \bil\ ($\kappa=0$) is computed.
The Kolmogorov-Smirnov statistic is then given by

\beq
\label{ks0}
D~
=~
\sqrt{ {\cal L} \int_0^{y_{\rm max}} dy ~ {d\sigma_1 \over dy} }
\quad
{\max}_{y}
\left|~
{\displaystyle
 \int_0^y             dy ~ {d\sigma_1 \over dy}
 \over
 \displaystyle
 \int_0^{y_{\rm max}} dy ~ {d\sigma_1 \over dy}}~
-~
{\displaystyle
 \int_0^y             dy ~ {d\sigma_0 \over dy}
 \over
 \displaystyle
 \int_0^{y_{\rm max}} dy ~ {d\sigma_0 \over dy}}~
\right|~
~,
\eeq

where the luminosity $\cal L$ is given by Eq.~(\ref{lum})
and the indices 0 and 1
refer to the cases $\kappa=0$ and $\kappa=1,$ respectively.
If $D=1.95$
the observed $\kappa=1$ distribution
has a probability of 99.9\%\ 
to be different from a distribution
which would have been observed if $\kappa=0$.
Of course,
if the observed \bil\ is an ordinary vector,
the corresponding Kolmogorov-Smirnov statistic 
is obtained from Eq.~(\ref{ks0})
by interchanging the indices 0 and 1.

We have plotted in Fig.~\ref{fks}
the $D=1.95$ boundaries 
in the $(m_B,\lambda)$ plane
for different values of the \cm\ energy.
The sensitivity increases 
the further away the \bil s are tested from threshold.
Far from threshold
the sensitivity is slightly better 
if the observed \bil s are ordinary vectors
rather than gauge \bil s,
because of the larger event rates.
In this case 
the distinction between the two types of bileptons 
can be made for values of the coupling $\lambda$
as low as a few per mil.

\section{Conclusions}

We have investigated the potential 
of the \ee\ collision mode 
of a \lc\ 
for discovering and studying doubly-charged bileptons 
in a model-independent way. 
For this we have used the most general renormalizable 
lepton-bilepton and photon-\bil\ interactions 
for scalar and vector bileptons belonging to 
different representations of the $SU(2)_L$ gauge group,
and we have studied the $s$-channel process 
$e^-e^-\rightarrow\mu^-\mu^-$ 
taking into account the effects of \isr, beam spread and polarization.

As expected, 
the \lc\ sensitivity to bilepton properties is best
when running on the resonance $\sqrt{s}=m_B.$  
With the planned machine parameters 
couplings to fermions can be tested 
down to less than $10^{-4}$ at the resonance.
Far from the resonance
there is still an important radiative return to the bilepton resonance  
due to \isr,
which makes it possible to discover \bil s
for couplings
down to less than $10^{-3}$
without having to resort to a collider energy scan.

If a \bil\ is discovered,
its nature can be investigated with several methods. 
The simplest way to discriminate between scalar and vector bileptons 
is to make use of polarization,
since flipping the helicity of the incoming electron beams 
would increase or decrease the production cross sections differently. 
The knowledge of its spin 
is thus a free by-product 
of the discovery of a \bil.

Recognizing Yang-Mills bileptons 
from minimally coupled vectors 
can be done by detecting hard photons in the process
$\e^-e^-\rightarrow L^{--}_{2\mu}\gamma.$ 
Because the angular distributions are different
the gauge nature of a light bilepton 
can easily be tested for couplings 
as small as $10^{-2}$
or even smaller.

\section*{Acknowledgments}

We are thankful to Dirk Graudenz
for his careful reading of the manuscript.
M.R. thanks the Spanish Ministry of Education for a post-doctoral grant.
His work is supported by the CICYT under grant AEN-96-1718.

\clearpage

\end{document}